\documentclass[sigconf]{acmart}
\usepackage{colortbl}
\usepackage{stfloats}  
\usepackage{booktabs}
\usepackage{algorithm}
\usepackage{algorithmic}
\usepackage{xcolor}
\usepackage{subcaption}
\usepackage{multirow}
\usepackage{tikz}
\usepackage[most]{tcolorbox}
\usetikzlibrary{positioning, arrows.meta, shapes.geometric, fit,
  patterns, calc, decorations.markings, decorations.pathreplacing}

\AtBeginDocument{%
  }

\setcopyright{none}
\renewcommand\footnotetextcopyrightpermission[1]{}

\begin{document}
\raggedbottom

\title{Agentic CPU-GPU Scheduling for Heterogeneous AI Workloads}

\author{Tianxi Lu}
\affiliation{%
  \institution{Brown University}
  \city{Providence}
  \state{Rhode Island}
  \country{USA}
}
\email{tianxi_lu@brown.edu}

\author{Sherief Reda}
\affiliation{%
  \institution{Brown University}
  \city{Providence}
  \state{Rhode Island}
  \country{USA}
}
\email{sherief_reda@brown.edu}

\renewcommand{\shortauthors}{Lu and Reda}

\begin{abstract}
Agentic AI systems compose heterogeneous tool workloads on shared
GPU/CPU infrastructure, yet existing frameworks assign all GPU-capable
tools to the GPU by default.
We profile 19 AI tools across GPU and CPU and find that 11 are
GPU-preferred, 4 are ambiguous, 1 is CPU-preferred, 
due to PCIe transfer dominance, and 3 are device-neutral, establishing
that blanket GPU-first scheduling is suboptimal.
We formulate device scheduling as assigning each tool to one of three
options---immediate GPU execution, queued GPU execution, or CPU
offload---under a VRAM budget, and identify two runtime factors that
cause end-to-end latency to diverge from static profiles: GPU
utilization contention and VRAM capacity contention.
We present an agentic scheduler that pairs an LLM agent with an algorithmic runtime
monitor, where the monitor expands what the LLM can observe---via
running averages, symmetric reprobing, swap reprobing, and exploration
hints---without ever prescribing which mapping to adopt.
Across 13 scenarios spanning serial execution, parallel contention, and
memory-constrained execution, the agentic scheduler reaches the
brute-force optimal mapping in all 13 scenarios, matching the best
classical baseline on mapping accuracy while avoiding
bandit-style exploration over complete mappings, and outperforming
HEFT, StarPU, and the all-GPU policy while requiring zero offline
training.
\end{abstract}



\maketitle
\pagestyle{plain}

\section{Introduction}
\label{sec:intro}
Modern agentic AI systems organize computation as multi-step workflows
centered around large language models (LLMs), which invoke external tools
to accomplish complex tasks~\cite{langgraph, autogen, crewai}.
These tools span diverse functionality, including deep-learning inference,
vector search, data processing, and external APIs.
Across such workflows, tool executions form a directed pipeline whose
end-to-end latency critically depends on how individual tools are mapped
to underlying hardware.

Despite this diversity, existing orchestration frameworks such as
LangGraph and LangChain adopt a uniform GPU-first policy, placing all
GPU-capable tools on the GPU.
This raises a fundamental systems question:
\emph{how should heterogeneous tools be scheduled across CPU and GPU
to minimize end-to-end latency?}
The problem is non-trivial because (i) tools have heterogeneous
profiles---compute-bound models favor GPUs while I/O-bound tools
favor CPUs; (ii) concurrent GPU execution causes contention that
deviates from isolated profiles; and (iii) limited VRAM imposes a
hard capacity constraint that expands the decision space beyond
binary placement.

Existing scheduling approaches are insufficient to address these
challenges.
Static heuristics (e.g., HEFT~\cite{heft} or bandit-based methods
such as UCB1~\cite{ucb1}) rely on fixed or locally updated estimates
and fail to capture global interactions under contention.
Learning-based approaches such as reinforcement-learning device
placement~\cite{mirhoseini2017, mirhoseini2018, placeto, regal}
require substantial offline training and struggle to generalize
across workloads with diverse tool compositions.
Meanwhile, modern agentic systems already employ an LLM as a central
planner that decomposes tasks, selects tools, and determines execution
order.
This naturally extends to device scheduling, where the LLM can reason
about placement decisions based on per-tool performance characteristics.

However, LLM-based scheduling alone is insufficient, particularly in the
presence of contention and memory constraints.
To address this, we design an \emph{agentic scheduler} that combines an
LLM with a runtime monitor.
The LLM performs high-level reasoning over device mappings, while the
monitor provides dynamic feedback by measuring execution under
contention, probing alternative placements, and exposing information
beyond static profiles.
This co-design enables effective scheduling without requiring offline
training.

This paper makes the following contributions.
\begin{itemize}
  \item We formalize heterogeneous scheduling for agentic workflows as a DAG-based optimization problem, where each tool invocation is mapped to CPU or GPU under precedence and GPU memory constraints to minimize end-to-end latency.

  \item We profile 19 AI tools with real-world datasets across CPU and GPU, identifying four distinct scheduling zones and showing that the commonly used GPU-first policy is often suboptimal.

\item We design an agentic scheduler that integrates an LLM with a runtime monitor. The LLM determines the scheduling of invoked tools, while the monitor provides dynamic feedback to expand the LLM's observation space, enabling the scheduler to escape local optima without offline training.

  \item We evaluate our system on 13 scenarios against four baselines (HEFT, UCB1, StarPU, and all-GPU). Our approach reaches the brute-force-optimal mapping in all 13 scenarios, matching UCB1 on mapping accuracy while avoiding its exploration over complete mappings and outperforming HEFT, StarPU, and all-GPU.
\end{itemize}

The remainder of this paper is organized as follows.
Section~\ref{sec:background} motivates and formulates the scheduling problem.
Section~\ref{sec:design} presents the design of our agentic scheduler.
Section~\ref{sec:eval} evaluates the system against classical baselines.
Section~\ref{sec:related} discusses related work, Section~\ref{sec:discussion} discusses limitations, and Section~\ref{sec:conclusion} concludes.

\section{Motivation}
\label{sec:background}

\subsection{Problem Formulation}

We model an agentic workflow as a directed acyclic graph (DAG)
$G=(V,E)$, where each node $v\in V$ represents a tool invocation and
each edge $(u,v)\in E$ denotes a data dependency: task $v$ can start only
after task $u$ completes.

\begin{figure}[t!]
  \centering
  \includegraphics[width=\columnwidth]{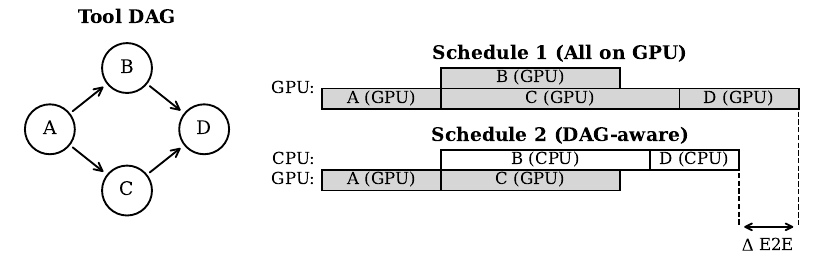}
  \caption{
Motivating example of heterogeneous placement for an agentic workflow.
End-to-end time is measured from the start of Tool A to the completion of
Tool D. A naive GPU-first policy may serialize execution under GPU
contention, while a dependency-aware placement improves parallelism and
reduces latency. The example shows CPU/GPU placement only; our full
formulation further distinguishes immediate GPU execution from queued GPU
execution under a VRAM budget.
}
  \label{fig:motivation}
\end{figure}

Each task is assigned to one of three execution modes:
\[
a_v \in \mathcal{A}
=
\{\texttt{cpu}, \texttt{gpu\_now}, \texttt{gpu\_queue}\}.
\]
The \texttt{cpu} mode executes the task on CPU and consumes no GPU
resources. The \texttt{gpu\_now} mode executes the task immediately on
GPU once its dependencies are satisfied, so its model and intermediate
state occupy GPU memory during execution. The \texttt{gpu\_queue} mode
also executes the task on GPU, but the task is deferred to a serialized
GPU queue. Queued tasks do not reserve VRAM while waiting; instead, they
pay an additional load-in or queuing overhead when they are later admitted
to the GPU.
Figure~\ref{fig:motivation} illustrates how different placements over
these modes can yield substantially different end-to-end latency for
the same DAG.

Let $\tau(v,a_v)$ denote the latency of task $v$ under execution mode
$a_v$, including any load-in or queuing overhead for
\texttt{gpu\_queue}. A placement $\pi:V\rightarrow\mathcal{A}$ is
feasible only if it respects the DAG dependencies and the GPU memory
budget. Specifically, for any time $t$, the total VRAM footprint of
tasks actively executing in \texttt{gpu\_now} mode cannot exceed the GPU
capacity $M$:
\[
\sum_{v\in \mathcal{G}_{\mathrm{now}}(t)} m(v) \le M,
\]
where $\mathcal{G}_{\mathrm{now}}(t)$ is the set of tasks assigned to
\texttt{gpu\_now} that are active at time $t$, and $m(v)$ is the VRAM
footprint of task $v$. Tasks assigned to \texttt{gpu\_queue} are admitted
sequentially and therefore do not contribute to this simultaneous
\texttt{gpu\_now} VRAM footprint while waiting.

The end-to-end latency of a workflow is the wall-clock time from workflow
launch to the completion of its final output. Let $C_v$ denote the
completion time of task $v$ under placement $\pi$ and schedule $S$. Since
workflow outputs correspond to sink nodes of the DAG, we define
\[
T_{\mathrm{e2e}}(G,\pi,S)
=
\max_{v\in \mathrm{Sink}(G)} C_v .
\]
The scheduling objective is therefore
\[
\min_{\pi,S} \; T_{\mathrm{e2e}}(G,\pi,S),
\]
subject to dependency feasibility and GPU-memory feasibility.

This formulation captures the central trade-off in agentic heterogeneous
execution. Assigning more tasks to \texttt{gpu\_now} may reduce isolated
task latency, but can increase GPU contention or exceed the VRAM budget.
Assigning a task to \texttt{gpu\_queue} preserves GPU execution without
reserving VRAM while waiting, but introduces load-in and queuing overhead.
Assigning a task to \texttt{cpu} avoids GPU pressure altogether, but may
be slower for GPU-preferred tools. Even without the \texttt{gpu\_queue}
option, this problem generalizes precedence-constrained heterogeneous
scheduling with resource-capacity constraints, which is known to be
NP-hard. The additional queued-GPU mode further expands the placement
space, motivating a scheduler that can reason about both static profiles
and runtime feedback.

\subsection{Workload Landscape}

The tools that comprise modern agentic pipelines span a wide range of
computational profiles.
At one end of the spectrum, deep-learning inference models such as
ResNet-50~\cite{resnet}, BERT~\cite{bert},
DistilBERT~\cite{distilbert}, Whisper~\cite{whisper},
CLIP~\cite{clip}, TrOCR~\cite{trocr}, and Llama~3.2~\cite{llama3} perform
heavily parallelizable operations (convolution, self-attention) that
achieve order-of-magnitude speedups on GPU.
At the other end, tools such as document parsers, log analyzers, and
HTTP API wrappers execute purely sequential or I/O-bound code that
derives no benefit from GPU parallelism.
Between these extremes lie tools whose optimal device depends on
runtime conditions: tabular aggregation kernels, image-resizing
pipelines, and vector-search routines that achieve modest GPU speedups
under isolation but may be penalized by contention when sharing the
GPU with heavier workloads.

We profile 19 such tools on an NVIDIA RTX~A5500 GPU paired with an
AMD EPYC~7713 CPU (Table~\ref{tab:tools}).
Based on the GPU speedup ratio $s = T_{\text{cpu}} / T_{\text{gpu}}$,
tools fall into four scheduling zones: GPU-preferred ($s > 5\times$,
11 tools), ambiguous ($0.5 \leq s \leq 5\times$, 4 tools),
CPU-preferred ($s < 0.5\times$, 1 tool whose GPU time is 97.5\% PCIe
transfer), and device-neutral ($s \approx 1\times$, 3 tools lacking
GPU implementations).
This distribution confirms that while the majority of tools do benefit
from GPU execution, a non-trivial fraction does not, and a blanket
GPU-first policy will misplace them.

\begin{table*}[t]
  \caption{The 19-tool library, profiled in isolation on NVIDIA
    RTX~A5500 + AMD EPYC~7713.  GPU and CPU execution times are the
    median of 30 runs.  VRAM is the peak GPU memory allocated during
    isolated execution.  Load is the one-time model load-in cost
    paid when a tool is launched via \emph{gpu\_queue}; the
    rightmost column reports the resulting single-shot speedup
    $\text{CPU}\,/\,(\text{GPU}+\text{Load})$, which collapses below
    $1\times$ for almost every tool once load cost is amortized over
    a single invocation.}
  \label{tab:tools}
  \vspace{-0.1in}
  \footnotesize
  \begin{tabular*}{\textwidth}{@{\extracolsep{\fill}}llrrrrrr@{}}
    \toprule
     & & \multicolumn{4}{c}{Isolated profiling} & \multicolumn{2}{c}{\emph{gpu\_queue}} \\
    \cmidrule(lr){3-6} \cmidrule(lr){7-8}
    Tool & Workload & GPU (ms) & CPU (ms) & Speedup & VRAM (MB) & Load (ms) & Speedup \\
    \midrule
        sentiment       & DistilBERT-SST2~\cite{distilbert,sst2}     & 7    & 450   & 64$\times$  & 316  & 1022  & 0.44$\times$    \\
    bert\_base      & BERT-base NLI~\cite{bert,mnli}             & 13   & 892   & 69$\times$  & 486  & 481   & 1.81$\times$    \\
    distilbert      & DistilBERT-MNLI~\cite{distilbert,mnli}     & 7    & 456   & 65$\times$  & 314  & 1079  & 0.42$\times$    \\
    reranker        & Cross-encoder~\cite{bert,sbert,msmarco}    & 8    & 440   & 55$\times$  & 170  & 739   & 0.59$\times$    \\
    minilm\_l6      & MiniLM-L6 embed~\cite{minilm,msmarco}      & 5    & 280   & 56$\times$  & 128  & 8890  & 0.03$\times$    \\
    llama\_3b       & Llama-3.2-3B gen.~\cite{llama3}            & 768  & 19465 & 25$\times$  & 6440 & 5278  & 3.22$\times$    \\
    resnet50        & ResNet-50 CNN~\cite{resnet,imagenet}       & 8    & 216   & 27$\times$  & 150  & 7699  & 0.03$\times$    \\
    clip            & CLIP ViT-B/32~\cite{clip}                 & 12   & 221   & 18$\times$  & 656  & 2944  & 0.07$\times$    \\
    whisper\_base   & Whisper-Base ASR~\cite{whisper,librispeech} & 35 & 473   & 14$\times$  & 442  & 1792  & 0.26$\times$    \\
    llama\_1b       & Llama-3.2-1B gen.~\cite{llama3}            & 447  & 8482  & 19$\times$  & 2484 & 2519  & 2.86$\times$    \\
    ocr             & TrOCR handwriting~\cite{trocr}             & 176  & 1386  & 8$\times$   & 248  & 3244  & 0.40$\times$    \\
    \midrule
    groupby         & Scatter-add agg.                           & 2    & 9     & 5$\times$   & 61   & 183   & 0.05$\times$    \\
    image\_resize   & Bilinear resize                            & 4    & 17    & 4$\times$   & 340  & 4008  & $<$0.01$\times$ \\
    audio\_features & STFT + MFCC                                & 0.8  & 2     & 2$\times$   & 25   & 364   & $<$0.01$\times$ \\
    vector\_search  & Top-$k$ sim.~\cite{faiss,msmarco}          & 18   & 43    & 2$\times$   & 3106 & 14701 & $<$0.01$\times$ \\
    \midrule
    doc\_merger     & Embedding reduction                         & 39   & 14    & 0.4$\times$ & 692  & 1218  & 0.01$\times$    \\
    \midrule
    web\_search     & HTTP API call                               & 128  & 127   & 1.0$\times$ & --   & --    & --              \\
    log\_analyzer   & Apache log parse                            & 641  & 638   & 1.0$\times$ & --   & --    & --              \\
    pdf\_parser     & PDF text extract                            & 6    & 8     & 1.3$\times$ & --   & --    & --              \\
    \bottomrule
  \end{tabular*}
  \vspace{2pt}
  \footnotesize VRAM is the peak GPU memory measured during isolated
single-tool execution (model weights + activations + cuBLAS workspace);
concurrent execution may share workspace and produce a lower aggregate
footprint than the per-tool sum.  ``--'' denotes the three neutral
tools (last three rows), which have no GPU-specific implementation.
\end{table*}

\subsection{Runtime Factors that Alter the Static Profile}
\label{sec:factors}

In a real agentic workflow, tools rarely execute in isolation.
Two orthogonal runtime factors cause end-to-end latency to diverge
from isolated-profile predictions: GPU utilization contention and
VRAM capacity contention.

\paragraph{Factor A: GPU utilization contention.}
Concurrent GPU tools compete for streaming multiprocessors, memory
bandwidth, and dispatch queues; per-tool latency can exceed the
isolated GPU time by a factor of two or more, while CPU latency is
unaffected.  Table~\ref{tab:factorA} shows this in scenario S5:
placing four GPU-preferred tools on GPU concurrently inflates
per-tool latency by 43--250\% above isolated profiles, and
end-to-end latency reaches $2.7\times$ the isolated upper bound.
The oracle mapping offloads \texttt{vector\_search\_x15} to CPU:
\texttt{vector\_search\_x15} itself becomes 78\% slower, but the three
remaining tools gain 18--44\% relief and end-to-end latency falls
from 988 to 812~ms.  The correct decision follows the contention
structure, not any one tool's isolated profile. Here, Oracle denotes an offline brute-force reference: the feasible
mapping with the lowest measured end-to-end latency among all mappings
enumerated by exhaustive search.

\begin{table}[t]
  \centering
  \small
  \caption{Per-tool latency (ms) in scenario S5. Isolated is the
    profiled single-tool time; All-GPU places all four tools on
    GPU concurrently; Oracle offloads \texttt{vector\_search\_x15}
    to CPU.  $\Delta_1$ is All-GPU vs.\ Isolated; $\Delta_2$ is Oracle vs.\ All-GPU.  Workloads are scenario-amplified; numbers are
    not directly comparable to Table~\ref{tab:tools}.}
  \label{tab:factorA}
  \vspace{-0.1in}
  \begin{tabular*}{\columnwidth}{@{}l@{\extracolsep{\fill}}rrrrr@{}}
    \toprule
    Tool
      & \shortstack{Isol.\\(ms)}
      & \shortstack{All-GPU\\(ms)}
      & $\Delta_1$
      & \shortstack{Oracle\\(ms)}
      & $\Delta_2$ \\
    \midrule
    ocr\_x2            & 370 & 988 & $+167\%$ & 812       & $-18\%$ \\
    whisper\_x4        & 136 & 270 & $+99\%$  & 171       & $-37\%$ \\
    vector\_search\_x15 & 277 & 397 & $+43\%$  & 706$^{*}$ & $+78\%$ \\
    image\_resize\_x30  & 145 & 508 & $+250\%$ & 286       & $-44\%$ \\
    \midrule
    E2E            & --  & 988 & --       & 812       & $-18\%$ \\
    \bottomrule
  \end{tabular*}
  \vspace{-0.1in}
  \footnotesize $^{*}$Runs on CPU under oracle mapping.
\end{table}

\paragraph{Factor B: VRAM capacity contention.}
When the aggregate memory footprint of concurrent GPU tools exceeds
available VRAM, the all-GPU mapping is infeasible rather than
merely suboptimal---the later-loading tool triggers out-of-memory.
A third option becomes useful: \emph{gpu\_queue}, in which a tool
defers execution until the concurrent phase completes and then
runs alone on the now-free GPU.  The Load column of
Table~\ref{tab:tools} shows gpu\_queue is not free: for a tool
invoked once per DAG, the load cost typically dominates, pushing
the single-shot speedup below $1\times$ for all but one tool.  The
scheduler must therefore choose gpu\_queue or cpu on a tool-by-tool
basis.

Each factor alone is sufficient to invalidate a static mapping, and
neither is visible to a scheduler that sees only isolated
profiles---motivating a scheduler that observes runtime feedback.

\section{Proposed Agentic Scheduler Design}
\label{sec:design}

\begin{figure*}[t]
  \centering
  \includegraphics[width=0.9\textwidth]{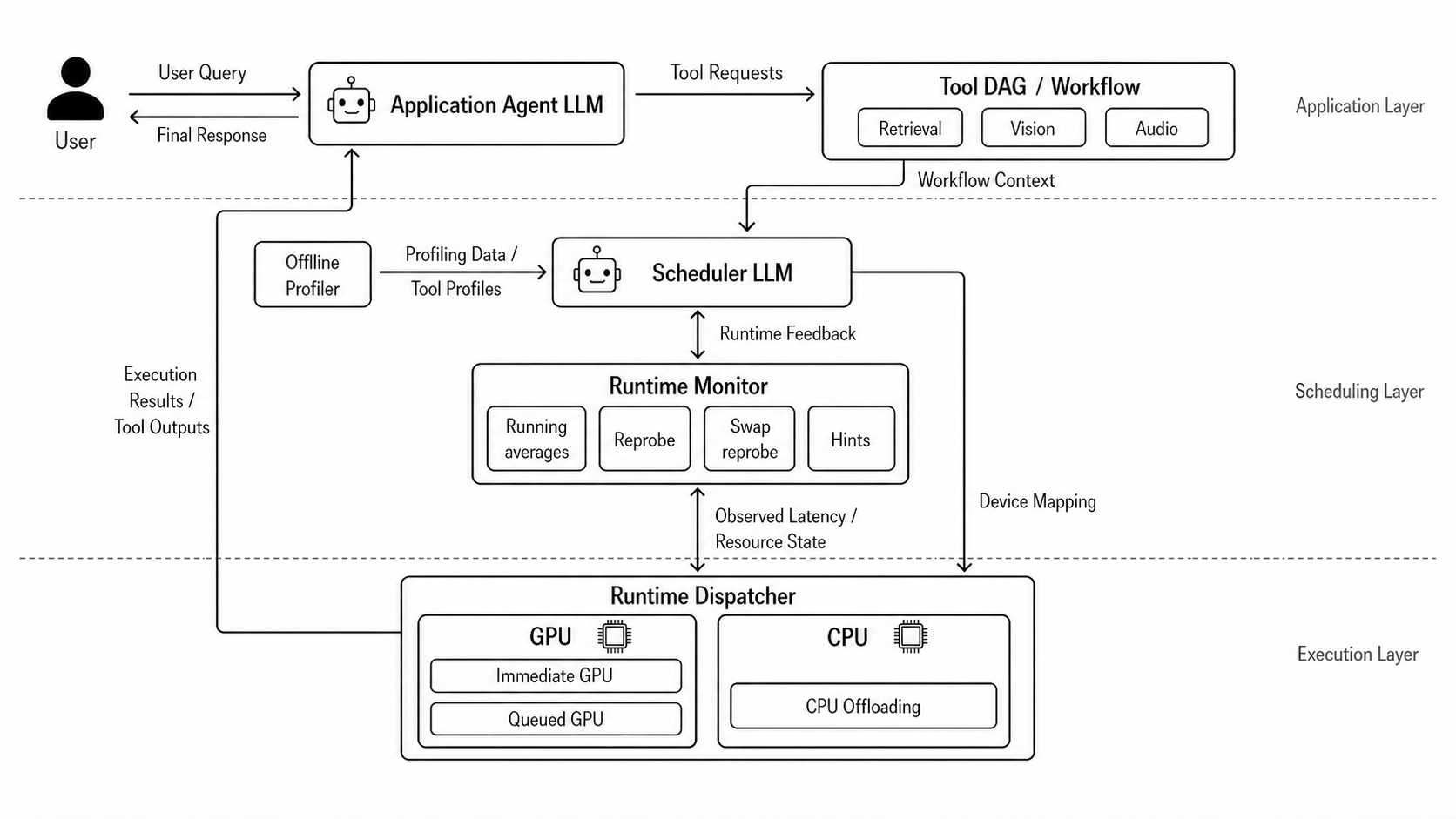}
  \vspace{-0.3in}
  \caption{System architecture. The offline profiler generates tool
  cards; the LLM scheduler selects a device mapping; the runtime
  monitor feeds back measured latency, feasibility, and bounded
  what-if probes for subsequent scheduling rounds.}
  \label{fig:overview}
\end{figure*}

The agentic scheduler comprises three components that interact in a
closed loop (Figure~\ref{fig:overview}).  The offline profiler
measures each tool's CPU latency, GPU latency, model load time, and
GPU memory footprint, and stores the results as structured tool
cards.  The LLM scheduler consumes these cards together with runtime
feedback and emits a device mapping for the next execution round.  The
runtime monitor observes execution outcomes, updates latency estimates
under the current contention regime, and, when memory pressure is
active, exposes bounded what-if measurements to the LLM.

The monitor is progressively activated.  In simple serial settings,
the profiling cards are usually sufficient and the monitor records
only execution history.  Under Factor~A, GPU utilization contention,
the monitor replaces isolated latency estimates with running averages
measured under actual concurrent execution.  Under Factor~B, VRAM
capacity contention, it additionally runs bounded reprobes and swap
reprobes to reveal feasible alternatives involving
\texttt{gpu\_now}, \texttt{gpu\_queue}, and \texttt{cpu}.  Thus, the
system uses measurement only when static profiles become unreliable.

\subsection{Tool Profiling}
\label{sec:profiling}

Each of the 19 base tools is profiled in isolation, measuring CPU
execution time, GPU execution time, model load time, and GPU memory
footprint when applicable.  We run each tool 30 times per device and
report the median.  The tools execute on public datasets, including
ImageNet~\cite{imagenet}, SST-2~\cite{sst2}, MNLI~\cite{mnli},
LibriSpeech~\cite{librispeech}, and MS~MARCO~\cite{msmarco}, with a
synthetic fallback when a dataset is unavailable.  For fan-out and
co-located experiments, where base tools are too short to produce
stable contention effects, we construct amplified tools by repeating
the same computation $k$ times, e.g., \texttt{bert\_base\_x12}.  This
preserves the tool's computational profile while scaling latency and
memory pressure into the operating range needed to study runtime
contention.

\subsection{Runtime Monitor}
\label{sec:monitor}

The runtime monitor is the feedback component between DAG execution
and the LLM scheduler.  It does not compute a schedule.  Instead, it
turns the consequences of previous mappings into observations that
can be appended to the next scheduler prompt.  In each round, the
monitor receives the LLM-selected mapping, the measured execution
result, and its internal state; it updates latency statistics,
optionally runs bounded what-if probes, and emits prompt fields such
as running averages, reprobe outcomes, swap outcomes, and exploration
hints.

Algorithm~\ref{alg:monitor} summarizes the per-round procedure.
Internally, the monitor keeps a sliding window of per-tool latencies,
the best end-to-end latency seen so far, and counters for stale
measurements and repeated mappings.

\begin{algorithm}[t]
\caption{Runtime Monitor (per round)}
\label{alg:monitor}
\footnotesize
\begin{algorithmic}[1]
\REQUIRE mapping $\mathbf{d}$, result $(E,\mathbf{t},\text{status})$,
state $S$
\STATE Update sliding-window latency statistics with $\mathbf{t}$
\STATE Update $S.\text{best}$ if $E < S.E^*$
\IF{the mapping $\mathbf{d}$ changed}
  \FOR{each tool $i$}
    \FOR{each alternative device $a \neq d_i$}
      \STATE $r \gets \textsc{Reprobe}(\mathbf{d}, i, a)$
      \IF{$r.\text{status}=\textsc{exceed}$ \AND $a=\texttt{gpu\_now}$}
        \FOR{each tool $j$ currently on \texttt{gpu\_now}}
          \STATE $\textsc{SwapReprobe}(\mathbf{d},
            i\!\rightarrow\!\texttt{gpu\_now},
            j\!\rightarrow\!\texttt{gpu\_queue})$
        \ENDFOR
      \ENDIF
    \ENDFOR
  \ENDFOR
\ENDIF
\STATE Check exploration-hint conditions from $S$
\STATE Build next prompt fields from history, running averages,
reprobe results, swap results, and hints
\end{algorithmic}
\end{algorithm}

\paragraph{Observation feedback.}
The passive path, Lines~1--2, addresses Factor~A.  Because offline
profiles are measured in isolation, GPU latency estimates can become
optimistic when several tools execute concurrently.  The monitor
therefore keeps a sliding window of recent observations for each
tool--device pair and reports the corresponding running average to
the LLM.  These averages expose the latency actually induced by the
LLM's previous mappings, allowing the scheduler to react to contention
without an explicit analytical GPU-sharing model.  The monitor also
tracks unobserved tool--device pairs so that stale estimates can be
reprobed when they may hide a profitable device change.

\paragraph{Bounded active exploration.}
The active path, Lines~3--14, is enabled under Factor~B.  Under Factor~B, the monitor operates over the three execution modes
defined in Section~\ref{sec:background}: \texttt{gpu\_now},
\texttt{gpu\_queue}, and \texttt{cpu}. When the all-\texttt{gpu\_now}
placement exceeds the VRAM budget, the scheduler must decide which tools
should remain resident on GPU, which should be deferred through the GPU
queue, and which should fall back to CPU.  Observation feedback alone may miss alternatives that
the current mapping never executes, so the monitor performs
single-tool reprobes by moving one tool to an alternative placement
while keeping the rest of the mapping fixed.  Feasible reprobes are
executed and reported as measured what-if evidence.

Single-tool moves are insufficient when a tool cannot move to
\texttt{gpu\_now} without displacing another resident tool.  In this
case, the monitor runs swap reprobes: the blocked tool is moved to
\texttt{gpu\_now}, while one current \texttt{gpu\_now} tool is moved
to \texttt{gpu\_queue}.  This exposes useful two-tool rearrangements
without enumerating the full $3^n$ mapping space.  Finally, if the LLM
repeats a stagnant mapping, the monitor may issue a controlled
exploration hint.  Such hints relax the search behavior but do not
specify a target mapping.

\subsection{LLM Scheduler}
\label{sec:llm}

We use an LLM as a flexible policy layer over heterogeneous scheduling
evidence.  The scheduler must reason jointly about DAG structure,
tool semantics, CPU/GPU profiles, memory footprints, cold-start load
costs, execution history, and monitor-generated what-if measurements.
Encoding all of these signals into a single rigid cost model is
brittle when workloads or contention regimes change.  By contrast,
the LLM consumes the same evidence as structured text and adapts its
mapping decision without additional training; it can also consume
softer guidelines such as ratio-based offload heuristics without
treating them as hard constraints, retaining the freedom to override
them when monitor evidence contradicts the rule.

The LLM's second strength is its ability to extract useful priors
from whatever evidence the prompt provides---profile cards when they
are present, semantic descriptions when they are not---and to revise
those priors as monitor feedback accumulates.  

We illustrate both flavors using scenario~S5, a fan-out workload
whose brute-force oracle places three tools on GPU and offloads
\texttt{vector\_search\_x15} to CPU;
Table~\ref{tab:factorA} reports
the resulting per-tool latencies under all-GPU and oracle mappings.

\paragraph{With profile data: oracle from prompt alone.}
When the prompt includes isolated profile cards, the LLM applies a
ratio test on each tool against an offload guideline and emits the
oracle mapping in round~0 (Figure~\ref{lst:s5_warm}).  It identifies
that \texttt{vector\_search\_x15}'s CPU time (387\,ms) is close to the
slowest GPU bottleneck while \texttt{image\_resize\_x30}'s CPU time
(521\,ms) is well above it, and reaches a 4/4 oracle match without
any monitor feedback.

\begin{figure}[t]
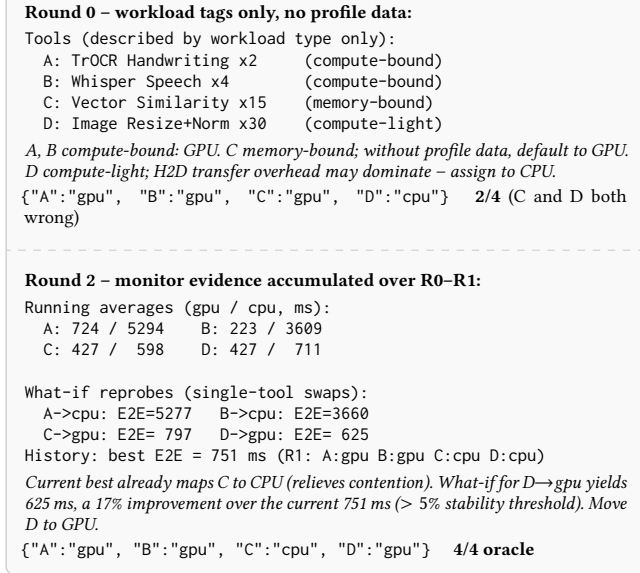

\footnotesize
\begin{tcolorbox}[colback=gray!5, colframe=gray!40, boxrule=0.5pt,
  left=4pt, right=4pt, top=3pt, bottom=3pt]
\textbf{Round 0 -- workload tags only, no profile data:}
\begin{verbatim}
Tools (described by workload type only):
  A: TrOCR Handwriting x2     (compute-bound)
  B: Whisper Speech x4        (compute-bound)
  C: Vector Similarity x15    (memory-bound)
  D: Image Resize+Norm x30    (compute-light)
\end{verbatim}
\emph{A, B compute-bound: GPU.  C memory-bound; without profile data,
default to GPU.  D compute-light; H2D transfer overhead may dominate
-- assign to CPU.}

\smallskip
\texttt{\{"A":"gpu", "B":"gpu", "C":"gpu", "D":"cpu"\}}
\quad\textbf{2/4} (C and D both wrong)

\tcblower
\textbf{Round 2 -- monitor evidence accumulated over R0--R1:}
\begin{verbatim}
Running averages (gpu / cpu, ms):
  A: 724 / 5294    B: 223 / 3609
  C: 427 /  598    D: 427 /  711

What-if reprobes (single-tool swaps):
  A->cpu: E2E=5277   B->cpu: E2E=3660
  C->gpu: E2E= 797   D->gpu: E2E= 625
History: best E2E = 751 ms (R1: A:gpu B:gpu C:cpu D:cpu)
\end{verbatim}
\emph{Current best already maps C to CPU (relieves contention).
What-if for D$\to$gpu yields 625\,ms, a 17\% improvement over the
current 751\,ms ($> 5\%$ stability threshold).  Move D to GPU.}

\smallskip
\texttt{\{"A":"gpu", "B":"gpu", "C":"cpu", "D":"gpu"\}}
\quad\textbf{4/4 oracle}
\end{tcolorbox}
\vspace{-0.1in}
\caption{Cold-start trajectory for scenario~S5.  Round~0: with no
profile data the LLM reasons from workload-type tags and incorrectly
keeps \texttt{vector\_search} on GPU while offloading
\texttt{image\_resize}.  Round~2: after the monitor accumulates
running averages and what-if reprobes, the entry
\texttt{D$\rightarrow$gpu = 625\,ms} drives the LLM to flip D back to
GPU, reaching the oracle mapping.}
\label{lst:s5_cold}
\end{figure}

\paragraph{Without profile data: monitor feedback corrects a wrong
semantic prior.}
When the prompt omits profile data, round~0 contains only
workload-type tags (\emph{compute-bound}, \emph{memory-bound},
\emph{compute-light}) and short tool descriptions, and the LLM falls
back to these semantic priors.  It places \texttt{vector\_search} on
GPU (no signal that its CPU time is small) and offloads
\texttt{image\_resize} to CPU (mistaking compute-light for
CPU-friendly), yielding a 2/4 oracle match (Figure~\ref{lst:s5_cold},
top).  Over the next two rounds the monitor accumulates per-tool
running averages and what-if reprobes for unobserved placements; by
round~3 the prompt contains a what-if entry
\texttt{D$\rightarrow$gpu: 625\,ms} that is shorter than the current
best E2E (751\,ms at round~1) and exceeds the prompt's 5\% stability
threshold.  The LLM cites this measurement directly to flip D back to
GPU, reaching the 4/4 oracle mapping (Figure~\ref{lst:s5_cold},
bottom).  The monitor does not prescribe a placement; it expands the
LLM's observation space so that combinatorially distant alternatives
become measured what-if evidence.

By contrast, a purely exploratory bandit baseline must rediscover
even obvious device preferences from scratch, while a static
heuristic cannot revise its assumptions after contention appears.
The LLM bridges these regimes: it consumes profile cards or semantic
descriptions as priors, and consumes monitor outputs as in-context
evidence for revision.

\begin{figure}[t]
\footnotesize
\begin{tcolorbox}[colback=gray!5, colframe=gray!40, boxrule=0.5pt,
  left=4pt, right=4pt, top=3pt, bottom=3pt]
\begin{verbatim}
Tool profiles (isolated):
  A TrOCR x2:           gpu=370  cpu=127654  ratio=345
  B Whisper x4:         gpu=136  cpu=  9305  ratio= 69
  C Vector search x15:  gpu=277  cpu=   739  ratio=2.7
  D Image resize x30:   gpu=145  cpu=  2201  ratio= 15

Offload guideline (ratio = cpu_time / gpu_time):
  ratio<1: CPU-preferred. 1-7: ambiguous, offload only
  if cpu_time < slowest GPU bottleneck under contention.
  ratio>=7: keep on GPU.
\end{verbatim}
\emph{A, B, D have ratios well above 7: keep on GPU.  C is ambiguous
(ratio 2.7); under contention the GPU bottleneck inflates above C's
isolated 277\,ms, so C's CPU time of 739\,ms can hide behind it while
freeing GPU bandwidth for A, B, D.  Offload C.}

\medskip
\texttt{\{"A":"gpu", "B":"gpu", "C":"cpu", "D":"gpu"\}}
\quad\textbf{4/4 oracle}
\end{tcolorbox}
\vspace{-0.1in}
\caption{Warm-start round~0 of scenario~S5 (chain-of-thought abridged
from the model output).  Profile cards plus the ratio guideline are
sufficient: the LLM identifies \texttt{vector\_search\_x15} as the
offload target and emits the oracle mapping without monitor feedback.}
\label{lst:s5_warm}
\end{figure}

The scheduler follows a standard agentic pattern: profiling cards act
as the knowledge base, runtime observations provide execution
context, and the scheduler prompt acts as the instruction layer.  The
LLM is called once per round and emits a JSON device mapping that is
reused for all DAG executions within that round.  In all experiments,
we use DeepSeek-R1-Distill-Qwen-32B-AWQ~\cite{deepseek_r1, awq}
served by vLLM~\cite{vllm}.  The scheduling model runs on GPUs
dedicated to scheduling, separate from the tool execution GPU, so LLM
inference does not interfere with measured tool latency.

\paragraph{Prompt structure.}
The prompt contains four information streams: (i) DAG topology and
execution semantics; (ii) per-tool profiling cards, including
isolated CPU/GPU times, load times, and memory footprints; (iii)
execution history, including end-to-end latency and feasibility
status for prior rounds; and (iv) monitor-generated what-if data,
including reprobes and swap outcomes when Factor~B activates Level~2
exploration.

The exact form of streams~(ii) and~(iv) is regime-specific.  For
binary CPU/GPU placement (S1--S10), the prompt includes ratio-based
offload guidelines as soft heuristics: \texttt{ratio}\,$<1$ is
CPU-preferred, $1{\le}\texttt{ratio}{<}7$ is ambiguous (offload only
if CPU time stays below the slowest GPU bottleneck under contention),
and \texttt{ratio}\,$\ge 7$ keeps the tool on GPU; the cutoff at 7
falls inside the empirical $5\times$--$8\times$ gap between ambiguous
and strong-GPU tools in Table~\ref{tab:tools}.  The LLM may override
these rules when monitor evidence contradicts them, as illustrated by
the round-3 flip of D in Figure~\ref{lst:s5_cold}.  For three-way
placement (S11--S13), the prompt replaces ratio thresholds with
structural rules---memory-budget feasibility, the \texttt{gpu\_queue}
load-cost trade-off, and the two-phase execution timeline---because
the dominant constraint is memory feasibility rather than per-tool
speedup.  In neither regime does the prompt prescribe a target
mapping; the rules describe relationships among tools, and the LLM
must compose them.

\section{Evaluation}
\label{sec:eval}

We evaluate whether an LLM-based scheduler, when coupled with
measurement feedback, can find high-quality heterogeneous mappings
across increasingly difficult scheduling regimes.  The evaluation is
organized around three questions:
(i) whether the scheduler can recover static CPU/GPU placement
decisions when isolated profiles are reliable;
(ii) whether monitor feedback corrects static profiles under GPU
utilization contention; and
(iii) whether bounded reprobes and swap reprobes enable the scheduler
to reach oracle mappings under explicit VRAM constraints.

\subsection{Experimental Setup}
\label{sec:setup}

All experiments are conducted on a server equipped with NVIDIA
RTX~A5500 GPUs and AMD EPYC~7713 CPU cores.  In every scenario, one
GPU is dedicated to the LLM scheduler and a separate GPU is used for
tool execution, so LLM inference does not interfere with measured
tool latency.  For VRAM-constrained scenarios (S11--S13), we impose
a manual VRAM ceiling on the tool execution GPU to simulate memory
pressure that would arise on smaller-VRAM devices.  All methods
compared within a scenario are executed under the same hardware
configuration.

We compare the proposed agentic scheduler against four baselines.
\emph{All-GPU} assigns every GPU-capable tool to the GPU, matching the
default policy used by many orchestration frameworks.  \emph{HEFT}
applies precedence-aware list scheduling using profiled task
costs~\cite{heft}.  \emph{UCB1} treats mapping selection as a bandit
optimization problem and chooses mappings via upper confidence
bounds~\cite{ucb1}.  \emph{StarPU} is implemented as a greedy runtime
heuristic that assigns tools according to estimated completion time
with online performance-model updates~\cite{starpu}.  For
memory-constrained scenarios, we additionally compare against a
memory-aware greedy baseline, since the above baselines do not
directly support three-way placement with explicit VRAM constraints.
Because each evaluation scenario contains four tools, the feasible
mapping space is small enough to enumerate: $2^4$ mappings for
S1--S10 and $3^4$ mappings for S11--S13 before feasibility filtering.
We therefore define the oracle as the feasible mapping with the lowest
measured median end-to-end latency under the same execution environment.
The oracle is an offline brute-force reference, not an online scheduling
method.

The agentic scheduler is evaluated in two modes.  In \emph{cold
start}, the LLM receives only tool names and the workflow topology,
with no profiling data; it must reason from semantic priors alone in
the first round.  In \emph{warm start}, the LLM is initialized with
isolated profile cards (CPU/GPU latency, model load time, VRAM
footprint) measured during the offline profiling phase.  Both modes
share the same monitor feedback loop in subsequent rounds.  Each mode
runs for 20 rounds.  In each round, the target workflow is executed
multiple times, and we report the median end-to-end latency.

We construct 13 four-tool evaluation scenarios that progressively
stress three scheduling regimes: precedence-constrained placement,
GPU-utilization contention, and VRAM-constrained three-way placement.
Table~\ref{tab:scenario-configs} lists the per-scenario tool
composition, decision space, memory budget, and oracle mapping;
\textsection\ref{sec:diamond}--\ref{sec:colocated} discuss the design
of each family in turn.

\begin{table*}[t]
  \caption{Per-scenario configuration summary.  Memory budget is
    denoted ``--'' when unconstrained (binary CPU/GPU placement is
    feasible under the device budget).  Total memory is the sum of
    per-tool VRAM footprints; values are rounded to the nearest MB.
    Oracle mappings are reported in $(A,B,C,D)$ order.}
  \label{tab:scenario-configs}
  \footnotesize
  \centering
  \vspace{-0.15in}
  \begin{tabular*}{\textwidth}{@{\extracolsep{\fill}}llllllcrl@{}}
    \toprule
    ID  & Tool A & Tool B & Tool C & Tool D
        & Decision & Budget & Total & Oracle \\
        &        &        &        &        & space & (MB) & (MB) & ($A$,$B$,$C$,$D$) \\
    \midrule
    \multicolumn{9}{@{}l}{\emph{Topology: $A\!\to\!(B\|C)\!\to\!D$}} \\
    S1  & whisper\_base & llama\_1b
        & sentiment     & web\_search  & GPU/CPU & --   & 3242 & GPU,GPU,GPU,CPU \\
    S2  & minilm\_l6    & doc\_merger
        & groupby       & reranker     & GPU/CPU & --   & 1051 & GPU,CPU,CPU,GPU \\
    S3  & whisper\_base & distilbert
        & doc\_merger   & web\_search  & GPU/CPU & --   & 1448 & GPU,GPU,CPU,GPU \\
    S4  & minilm\_l6    & vector\_search
        & doc\_merger   & distilbert   & GPU/CPU & --   & 4240 & GPU,GPU,CPU,GPU \\

    \midrule
    \multicolumn{9}{@{}l}{\emph{Topology: $A\|B\|C\|D$}} \\
    S5  & ocr\_x2        & whisper\_x4      & vector\_search\_x15 & image\_resize\_x30 & GPU/CPU & -- & 4136 & GPU,GPU,CPU,GPU \\
    S6  & ocr\_x2        & resnet50\_x15    & groupby\_x60        & vector\_search\_x15 & GPU/CPU & -- & 3565 & GPU,GPU,GPU,CPU \\
    S7  & ocr\_x2        & clip\_x12        & bert\_base\_x12     & whisper\_x4         & GPU/CPU & -- & 1832 & GPU,GPU,GPU,GPU \\
    S8  & ocr\_x2        & resnet50\_x15    & vector\_search\_x15 & groupby\_x60        & GPU/CPU & -- & 3565 & GPU,GPU,CPU,GPU \\
    S9  & whisper\_x4    & clip\_x12        & resnet50\_x15       & image\_resize\_x30  & GPU/CPU & -- & 1588 & GPU,GPU,GPU,GPU \\
    S10 & ocr\_x2        & bert\_base\_x12  & groupby\_x60        & image\_resize\_x30  & GPU/CPU & -- & 1135 & GPU,GPU,GPU,CPU \\

    \midrule
    \multicolumn{9}{@{}l}{\emph{Topology: $A\|B\|C\|D$ with VRAM budget}} \\
    S11 & ocr\_x2        & clip\_x12        & bert\_base\_x12 & whisper\_x4 & GN/GQ/CPU & 1700 & 1832 & GN,GN,GQ,GN \\
    S12 & ocr\_x2        & bert\_base\_x12  & groupby\_x60    & image\_resize\_x30 & GN/GQ/CPU & 1700 & 1135 & GN,GN,GN,CPU \\
    S13 & clip\_x12      & bert\_base\_x12  & whisper\_x4     & groupby\_x60 & GN/GQ/CPU & 1100 & 1645 & GN,GQ,GN,CPU \\
    \bottomrule
  \end{tabular*}
  \vspace{-0.1in}
\end{table*}

\subsection{Precedence-Constrained Placement (S1--S4)}
\label{sec:diamond}

Scenarios S1--S4 use the diamond workflow
$A\!\to\!(B\parallel C)\!\to\!D$ to study heterogeneous placement
under precedence constraints alone.  The aggregate memory footprint
fits within device capacity and at most two tools share the GPU
concurrently, so neither runtime factor is active; the scheduling
problem reduces to per-tool device selection given DAG position, and
isolated profiles provide an adequate decision signal.  The four
scenarios are constructed to span the qualitative tool categories of
Table~\ref{tab:tools}---strong-GPU, ambiguous, CPU-preferred, and
device-neutral---so that no uniform device policy recovers the oracle
mapping (Table~\ref{tab:scenario-configs}) without per-tool reasoning.

S1 combines two strong-GPU tools (\texttt{whisper\_base},
\texttt{llama\_1b}) with a lightweight classifier (\texttt{sentiment})
and a neutral API call (\texttt{web\_search}).  The parallel branch is
highly asymmetric ($457$\,ms versus $6$\,ms isolated GPU time), and
the neutral tool tests whether the scheduler refrains from offloading
workloads that derive no benefit from either device.  S2 is the
principal counter-heuristic case: the parallel branch places
\texttt{doc\_merger}, the only CPU-preferred tool in our library,
alongside the ambiguous \texttt{groupby}; both are short
embedding-style operators with similar workload-class descriptions,
yet their optimal devices diverge---a separation that priors based on
workload semantics alone cannot resolve.  S3 draws one tool from each
scheduling zone and asks the scheduler to identify the single
profitable offload while retaining the strong-GPU and ambiguous tools
on the GPU.  S4 places two superficially similar offload candidates in
the parallel branch, the ambiguous \texttt{vector\_search} (isolated
speedup $1.7\times$) and the CPU-preferred \texttt{doc\_merger}
($0.4\times$); only the latter benefits from offloading, and a
ratio-threshold rule applied uncritically would offload both.

\begin{figure*}[t]
  \centering
  \includegraphics[width=\textwidth]{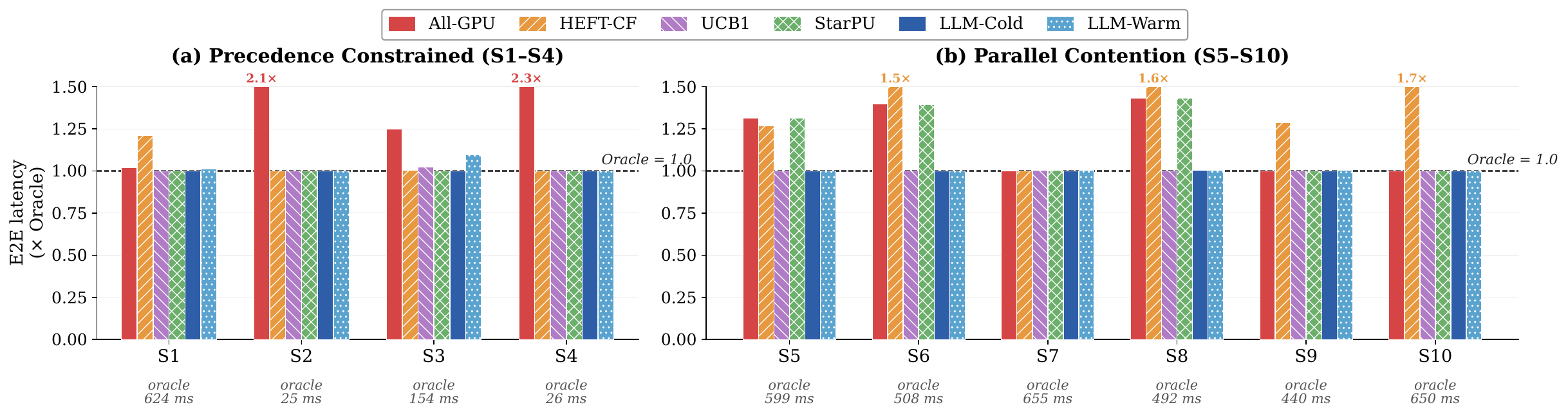}
  \caption{End-to-end latency normalized to the brute-force oracle
  for S1--S10, the feasible lower bound by construction.
  (a)~Precedence-constrained scenarios; (b)~fan-out scenarios with
  parallel contention.  Adaptive strategies (UCB1, StarPU, LLM-Cold,
  LLM-Warm) report the best of 20 rounds; bars taller than
  $1.5\times$ are annotated with their actual ratio.  Single-round
  ratios slightly below $1.0$ reflect run-to-run measurement noise
  against the oracle reference, not improvement over the oracle.}
  \label{fig:main_results}
  \vspace{-0.15in}
\end{figure*}

Figure~\ref{fig:main_results}(a) reports the best end-to-end latency
achieved by each strategy.  On the two high-gain scenarios, S2 and
S4, the agentic scheduler reaches oracle-level latency in both cold
and warm modes, achieving $2.60\times$ and $2.83\times$ speedup over
All-GPU, respectively.  Classical baselines also perform well in this
regime: HEFT, UCB1, and StarPU generally recover the correct
offloading decisions because the static profiles are reliable.  The
LLM's contribution in S1--S4 is therefore not superiority over
classical scheduling heuristics, but evidence that LLM-based
scheduling can recover profile-driven heterogeneous mappings without a
domain-specific scheduling rule.

Table~\ref{tab:oracle_match} reports oracle-mapping accuracy for
S1--S10.  Under precedence-constrained execution, UCB1 and the LLM
reach 4/4 placement matches on all four scenarios; HEFT-CF reaches
4/4 only on S4 and StarPU misses one placement on S2.  UCB1 requires
multiple exploration rounds to obtain these mappings, whereas the
LLM reaches the same mapping quality from the profiling cards in the
first scheduling round.

\begin{table}[t]
  \caption{Oracle-mapping accuracy under precedence-constrained
    placement (S1--S4) and GPU-utilization contention (S5--S10).  Each
    entry reports the best match, out of four tool placements, achieved
    over 20 rounds.  Bold entries indicate 4/4 oracle matches.}
  \label{tab:oracle_match}
  \vspace{-0.05in}
  \footnotesize
  \begin{tabular*}{\columnwidth}{@{\extracolsep{\fill}}lccccc@{}}
    \toprule
    & \multicolumn{4}{c}{Diamond DAG} & \\
    \cmidrule(lr){2-5}
    Strategy    & S1  & S2  & S3  & S4 & $\Sigma$\,4/4 \\
    \midrule
    HEFT-CF~\cite{heft}  & 3/4 & 3/4 & 3/4 & \textbf{4/4} & 1 \\
    UCB1~\cite{ucb1}     & \textbf{4/4} & \textbf{4/4} & \textbf{4/4} & \textbf{4/4} & 4 \\
    StarPU~\cite{starpu} & \textbf{4/4} & 3/4 & \textbf{4/4} & \textbf{4/4} & 3 \\
    LLM                  & \textbf{4/4} & \textbf{4/4} & \textbf{4/4} & \textbf{4/4} & 4 \\
    \bottomrule
  \end{tabular*}
  \vspace{0.05in}
  \begin{tabular*}{\columnwidth}{@{\extracolsep{\fill}}lccccccc@{}}
    \toprule
    & \multicolumn{6}{c}{Fan-out DAG} & \\
    \cmidrule(lr){2-7}
    Strategy    & S5  & S6  & S7  & S8  & S9  & S10 & $\Sigma$\,4/4 \\
    \midrule
    HEFT-CF~\cite{heft}  & 3/4 & 3/4 & \textbf{4/4} & 3/4 & 3/4 & 2/4 & 1 \\
    UCB1~\cite{ucb1}     & \textbf{4/4} & \textbf{4/4} & \textbf{4/4} & \textbf{4/4} & \textbf{4/4} & \textbf{4/4} & 6 \\
    StarPU~\cite{starpu} & 3/4 & 3/4 & \textbf{4/4} & 3/4 & \textbf{4/4} & \textbf{4/4} & 3 \\
    LLM                  & \textbf{4/4} & \textbf{4/4} & \textbf{4/4} & \textbf{4/4} & \textbf{4/4} & \textbf{4/4} & 6 \\
    \bottomrule
    \multicolumn{8}{r}{Overall 4/4: HEFT-CF 2/10, UCB1 10/10, StarPU 6/10, LLM 10/10.}
  \end{tabular*}
  \vspace{-0.05in}
\end{table}

\subsection{Contention-Aware Placement (S5--S10)}
\label{sec:fanout}

Scenarios S5--S10 use the fan-out topology
$A\parallel B\parallel C\parallel D$ to activate Factor~A, GPU
utilization contention.  All four amplified tools become ready
simultaneously and contend for GPU execution resources, while the
aggregate memory footprint remains within the VRAM budget so that
Factor~B is inactive; the decision space remains binary CPU/GPU.
Each scenario in this family is constructed to expose a distinct
contention pattern that defeats a different fixed policy.

S5 combines a heavy compute-bound anchor (\texttt{ocr\_x2}), a
strong-GPU tool (\texttt{whisper\_x4}), an ambiguous
\texttt{vector\_search\_x15} ($1.7\times$), and a compute-light
\texttt{image\_resize\_x30} ($3.7\times$); ratios are inherited from
the base tools in Table~\ref{tab:tools}.  The ambiguous tool is the
genuine offload target, while the compute-light tool is a decoy: its
higher isolated speedup invites offloading, but its CPU latency
exceeds the GPU bottleneck under contention.  S6 repeats the anchor pattern with a
milder workload mix; the parallel-branch contention is small enough
that an aggressive offload heuristic is penalized, testing whether the
scheduler refrains from offloading when contention does not warrant
it.  S7 places four tools that are all strong-GPU under isolation
(speedups $8$--$79\times$ in Table~\ref{tab:tools}, CPU latencies in
the seconds range), making all-GPU the oracle; this scenario tests
whether the scheduler resists over-offloading when contention is real
but every alternative is worse.  S8 places two ambiguous tools
(\texttt{vector\_search\_x15}, \texttt{groupby\_x60}) alongside two
strong-GPU anchors, asking the scheduler to discriminate between two
plausible offload candidates and pick the one whose CPU latency hides
behind the surviving GPU bottleneck.  S9 contains three tools with
similar isolated GPU times (no clear speedup cliff) plus a single
ambiguous candidate; the offloading decision is invisible from
isolated profiles alone and only becomes apparent once contention
measurements accumulate.  S10 stages the heaviest contention
configuration in this family: a strong-GPU anchor (\texttt{ocr\_x2})
dominates end-to-end latency, and the scheduler must identify the
ambiguous tool whose offload most relieves the anchor rather than the
ambiguous tool with the largest nominal speedup ratio.

Figure~\ref{fig:main_results}(b) and the lower panel of
Table~\ref{tab:oracle_match} show a sharper separation among
strategies.  HEFT-CF reaches 4/4 in only one of six scenarios because
its static contention factor cannot capture the actual interference
among four concurrent GPU tools; StarPU's online model helps in S9
and S10 but its greedy rule still misplaces in half of the fan-out
cases.  UCB1 and the LLM both reach 4/4 on all six, with different
mechanisms: UCB1 treats each complete mapping as an independent
bandit arm and explores over multiple rounds, whereas the LLM folds
monitor-measured running averages into each prompt and revises
placements directly, without per-mapping trials.

\subsection{Three-Way Placement under VRAM Constraints (S11--S13)}
\label{sec:colocated}

Scenarios S11--S13 evaluate the three-way placement setting with
explicit VRAM accounting.  Each tool may be assigned to
\texttt{gpu\_now}, \texttt{gpu\_queue}, or \texttt{cpu}:
\texttt{gpu\_now} tools execute immediately and count toward the
simultaneous VRAM budget; \texttt{gpu\_queue} tools execute on the GPU
after the immediate phase and do not count toward the same budget;
CPU tools avoid GPU memory pressure but may pay higher latency.  Two
scenarios in this family deliberately reuse tool sets from the
fan-out family under tighter memory budgets, isolating the effect of
memory pressure on otherwise contention-dominated workloads.
Per-scenario budgets, total footprints, and oracle mappings are listed
in Table~\ref{tab:scenario-configs}.

S11 reuses the four strong-GPU tools of S7 under a 1700\,MB budget
against a 1832\,MB total footprint, making the all-immediate-GPU
mapping marginally infeasible.  Because every tool is strongly
GPU-preferred (Table~\ref{tab:tools}), CPU offload is uniformly
expensive, and the oracle defers one tool to \texttt{gpu\_queue}
rather than to CPU; this scenario tests whether the scheduler
discovers that queue placement is preferable to CPU when GPU speedup
is large.  S12 reuses the tool set of S10 under the same 1700\,MB
budget but with a 1135\,MB total footprint, so the immediate-GPU
mapping remains feasible and CPU and queue alternatives are available
without being forced.  S12 serves as a control: the oracle here is
the same binary CPU/GPU mapping as in the unconstrained S10, and the
scenario tests whether the scheduler avoids over-reacting to memory
information when no memory pressure is present.  S13 combines tools
across scheduling zones under a tight 1100\,MB budget and a 1645\,MB
total, configured so that the oracle requires all three placement
modes simultaneously: two tools on \texttt{gpu\_now}, one on
\texttt{gpu\_queue}, and one on \texttt{cpu}.  The non-trivial
structure is that reaching the oracle requires a two-tool exchange
between the immediate-GPU set and the queue, which cannot be revealed
by any single local device change.

Table~\ref{tab:coloc} summarizes the results.  The memory-aware greedy
baseline assigns tools to \texttt{gpu\_now} by GPU speedup until the
budget is exhausted, then falls back to CPU.  This policy is safe but
locally greedy: it cannot discover that deferring one strongly
GPU-preferred tool to \texttt{gpu\_queue} may free enough VRAM for a
more valuable immediate GPU placement.  The agentic scheduler reaches
the oracle mapping in all three scenarios, converging by Round~1 in
S12, Round~2 in S11, and Round~3 in S13.

\begin{table*}[t]
  \caption{Three-way placement results under explicit VRAM accounting.
``Oracle-map E2E'' and ``LLM best-map E2E'' report measured latency as
$\mu \pm \sigma$ with sample size in parentheses.  In S13, the
best-found mapping differs from oracle only in off-critical-path
tool~D, whose CPU and \texttt{gpu\_queue} placements are within
run-to-run variance.}
  \label{tab:coloc}
  \small
  \centering
  \begin{tabular*}{\textwidth}{@{\extracolsep{\fill}}lrrrcccl@{}}
    \toprule
    Scenario & Budget & Total Mem. & Greedy E2E & Oracle-map E2E (ms) & LLM best-map E2E (ms) & Match & First 4/4 \\
    \midrule
    S11 & 1700 & 1832 & 6837 & $1715 \pm 24$ (15) & $1715 \pm 24$ (15) & \textbf{4/4} & R2 \\
    S12 & 1700 & 1135 &  845 & $\phantom{0}798 \pm 38$ (18) & $\phantom{0}798 \pm 38$ (18) & \textbf{4/4} & R1 \\
    S13 & 1100 & 1645 & 4685 & $1577 \pm \phantom{0}7$ \phantom{0}(4) & $1566 \pm 30$ (12) & \textbf{4/4} & R3 \\
    \bottomrule
  \end{tabular*}
\end{table*}

In S12, the LLM reaches oracle at Round~1 from reprobe data and
remains stable for most subsequent rounds; recognizing the
CPU-preferred tool is the only decision.  In S11, the LLM initially
selects a feasible local optimum by placing one tool on CPU; the
monitor's cpu-to-\texttt{gpu\_queue} reprobe reveals that queue
placement is faster than CPU execution, and swap reprobing then
exposes a two-tool exchange that reaches the oracle mapping at
Round~2.  S13 is the hardest of the three because reaching the
oracle requires both a queue placement and a non-obvious swap; we
analyze its convergence trajectory in detail below.

\subsection{Convergence Analysis on S13}
\label{sec:casestudy}

We analyze S13 to show how runtime timing observations drive convergence from an initially incorrect placement to the oracle mapping.
Scenario S13 places four amplified tools within a 1100\,MB VRAM
budget: \texttt{clip\_x12} (A, 656\,MB),
\texttt{bert\_base\_x12} (B, 486\,MB),
\texttt{whisper\_x4} (C, 442\,MB), and \texttt{groupby\_x60}
(D, 61\,MB).  The oracle mapping places A and C on
\texttt{gpu\_now}, B on \texttt{gpu\_queue}, and D on CPU.  This
mapping uses 1098\,MB in the immediate GPU phase, which satisfies the
1100\,MB budget, while deferring B to the GPU queue.  The key
difficulty is that B is strongly GPU-preferred, yet moving B out of
the immediate GPU set frees enough VRAM to place C on
\texttt{gpu\_now}.  This is a two-tool rearrangement that cannot be
revealed by a single local device change.

\begin{figure*}[t]
  \centering
  \includegraphics[width=0.8\textwidth]{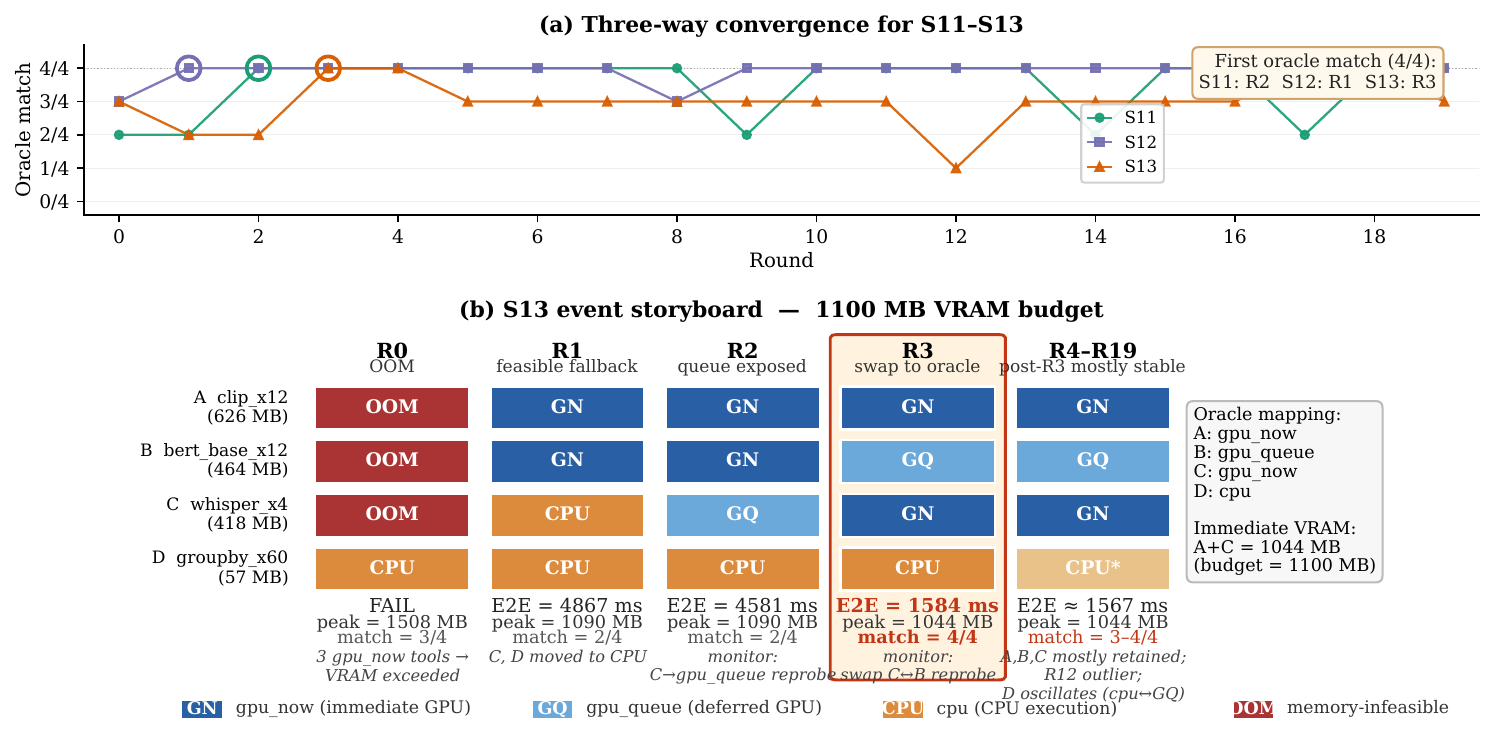}
  \vspace{-0.2in}
  \caption{S13 convergence trace with per-round E2E latency.  The monitor
first restores feasibility, then exposes C's \texttt{gpu\_queue}
alternative, and finally reveals the C$\leftrightarrow$B swap that
reaches the oracle mapping.}
  \label{fig:convergence}
  \vspace{-0.2in}
\end{figure*}

Figure~\ref{fig:convergence} summarizes the S13 convergence story in three phases.

In the first phase, the LLM repairs feasibility.  At Round~0, it
attempts to place the three GPU-preferred tools A, B, and C on
\texttt{gpu\_now}, which exceeds the 1100\,MB budget.  At Round~1, it
moves C and D to CPU, producing a feasible but slow mapping with
E2E $\approx$ 4867\,ms and a 2/4 oracle match.  This step shows that
the LLM can correct an infeasible initial mapping, but it has not yet
discovered the queue-based alternative.

In the second phase, the monitor exposes the value of
\texttt{gpu\_queue}.  Starting from the Round~1 mapping, the monitor
reprobes C on \texttt{gpu\_queue} and measures
E2E $\approx$ 4581\,ms.  The improvement is not yet sufficient to
reach oracle, but it gives the LLM direct evidence that C should not
remain on CPU.  The LLM adopts the queue placement at Round~2, which
still registers a 2/4 oracle match.

In the third phase, swap reprobing reveals the oracle rearrangement.
With C on \texttt{gpu\_queue}, a single-tool reprobe that moves C to
\texttt{gpu\_now} is infeasible because A, B, C, and D together exceed
the VRAM budget.  This failed reprobe triggers swap reprobing.  The
monitor tests the exchange C$\leftrightarrow$B, moving C to
\texttt{gpu\_now} and B to \texttt{gpu\_queue}.  The resulting
immediate GPU memory is A + C = 1098\,MB, which satisfies the budget,
and the measured E2E is approximately 1584\,ms.  The LLM adopts this
mapping at Round~3, reaching the oracle 4/4 placement.

After Round~3, the scheduler mostly retains the oracle placement for
A, B, and C, and D oscillates between \texttt{cpu} and
\texttt{gpu\_queue} with negligible effect because D is not on the
critical path.  One exploration outlier occurs at Round~12, where the
LLM transiently assigns A to \texttt{gpu\_queue} and D to
\texttt{gpu\_queue} (1/4 match, E2E $\approx$ 2571\,ms); the monitor
feedback redirects the scheduler back to the oracle placement in the
subsequent round.  We do not claim continuous 4/4 stability across
R4--R19; rather, the scheduler retains the oracle A/B/C placement in
most post-R3 rounds, the remaining variation is primarily D, and the
average E2E across oracle-aligned rounds with D on CPU is within
about 1\% of that with D on \texttt{gpu\_queue}.

Taken together, the evaluation shows that the LLM scheduler is
sufficient for profile-driven placement when static measurements are
trustworthy, while the monitor becomes essential as runtime effects
invalidate isolated profiles.  Running averages address utilization
contention, and bounded reprobes plus swap reprobes expose feasible
three-way placements under memory pressure.  The resulting system
reaches the oracle mapping in all 13 scenarios; small residual
deviations of measured end-to-end latency around the oracle reflect
run-to-run variation for latency-equivalent mappings rather than
improvement over the oracle.

\section{Discussion}
\label{sec:discussion}

The monitor is intended to expand the scheduler's observations rather
than encode a scheduling policy.  Symmetric reprobing fills missing
device measurements, swap reprobing tests a bounded set of two-tool
exchanges, and exploration hints trigger diversity when repeated
rounds stop producing new evidence.

The progressive design also keeps monitoring cost proportional to the
runtime factors present in the workload.  Serial execution needs no
monitoring, Factor~A requires only passive latency logging, and
Factor~B activates reprobes and exploration.  A deployment can
therefore begin with lightweight monitoring and enable the full monitor
only when scheduling quality stagnates under VRAM pressure.

The profiling landscape is hardware-dependent, but the scheduler
ports to new hardware by re-running the offline profiler rather than
retraining: the LLM reasons over the measured profiles in its prompt.

The current system provides no formal convergence guarantee.  In our
experiments, the monitor guides the LLM to the oracle mapping within
1--3 rounds, but the exploration policy is still heuristic: it expands
the observation space without proving that every starting point can
reach the global optimum.  A future data-driven policy could trigger
exploration only when runtime measurements reveal unexplored feasible
placements.

\section{Related Work}
\label{sec:related}

Heterogeneous task scheduling has a long history in the parallel
computing literature.
HEFT~\cite{heft} remains the canonical list-scheduling heuristic,
assigning tasks to the processor that minimizes earliest finish time
based on profiled task costs and communication delays.
StarPU~\cite{starpu} extends this approach with runtime performance
models that adapt to observed execution behavior.
System-level frameworks such as HeteroSpark~\cite{heterospark}
dispatch machine-learning tasks across GPU-accelerated clusters but
rely on fixed placement policies rather than online decisions.
These algorithms target binary processor placement (CPU or GPU)
without memory constraints, and they rely on domain-specific cost
models rather than general-purpose reasoning.
Our work extends the scheduling vocabulary to three device options
under explicit memory budgets and demonstrates that an LLM can
substitute for hand-crafted heuristics when paired with an
appropriate observation mechanism.

Reinforcement learning has been applied to device placement for
deep-learning computation graphs.
Mirhoseini et al.~\cite{mirhoseini2017, mirhoseini2018} train an
RL agent to assign TensorFlow operations across devices, an approach
subsequently generalized by Placeto~\cite{placeto} and
REGAL~\cite{regal}.
These methods require thousands of training episodes and target
operator-level granularity within a single model rather than
tool-level scheduling across heterogeneous workloads.
In contrast, the agentic scheduler requires zero training and
achieves comparable placement quality through in-context learning
from a small number of runtime observations.

Recent work has explored LLMs for systems optimization, including
compiler pass ordering~\cite{llm_compiler} and database
configuration tuning~\cite{llm_db_tuning}.
These applications share the premise that LLMs can reason about
system configurations from structured descriptions, but they operate
in single-shot or few-shot settings without the closed-loop
observation feedback that characterizes our monitor co-design.
To our knowledge, this work is the first to apply an LLM to
heterogeneous device scheduling for AI tool workloads and the first
to identify the need for a co-designed exploration engine that
compensates for the LLM's combinatorial reasoning limitations.

\section{Conclusion}
\label{sec:conclusion}

We presented an agentic CPU--GPU scheduler for heterogeneous AI tool
workloads.  The scheduler combines LLM-based placement decisions with
an algorithmic runtime monitor, allowing it to reason over static tool
profiles, execution history, and targeted what-if measurements without
offline training.  Our results show that static profiles are sufficient
only when runtime contention is absent: GPU-utilization contention
requires measured feedback, while VRAM capacity contention requires
bounded reprobes, swap discovery, and controlled exploration. Across 13 scenarios and four classical baselines, the scheduler reaches
the brute-force-optimal mapping while adapting to contention regimes
that defeat fixed-profile reasoning.  The central design principle is a
separation of observation and decision: the monitor expands the evidence
available to the LLM, while the LLM remains responsible for selecting
the mapping.  This co-design suggests a practical way to use LLMs in
systems optimization problems where the search space is combinatorial
but good decisions are possible once the right runtime evidence is made
explicit.

\bibliographystyle{ACM-Reference-Format}
\bibliography{references}

\clearpage

\appendix
\section*{Supplementary Material}
\addcontentsline{toc}{section}{Supplementary Material}

\section{S13 Ablation Study}
\label{app:ablation}

We ablate S13 to isolate the contribution of each monitor mechanism.
Table~\ref{tab:ablation} shows that removing any one mechanism
prevents the system from reaching the oracle mapping.  The failure
modes align with the three-step discovery chain discussed in the main
paper: cpu-to-queue reprobing is needed to expose C's
\texttt{gpu\_queue} alternative; swap reprobing is needed to expose
the C$\leftrightarrow$B exchange; and exploration hints are needed to
escape stagnant local optima when repeated mappings stop producing new
evidence.  These mechanisms are therefore
complementary rather than redundant.

\begin{table}[htbp]
  \caption{S13 ablation.  Each monitor mechanism is necessary for
  reaching the oracle mapping.  ``LLM only'' denotes the scheduler
  without runtime-monitor feedback.}
  \label{tab:ablation}
  \small
  \centering
  \begin{tabular*}{\columnwidth}{@{\extracolsep{\fill}}lcc@{}}
    \toprule
    Configuration & Best Match & Best E2E \\
    \midrule
    Full system                 & \textbf{4/4} & 1529\,ms \\
    w/o cpu$\to$queue reprobe   & 1/4          & 4853\,ms \\
    w/o swap reprobe            & 1/4          & 2174\,ms \\
    w/o exploration hints & 1/4    & 5060\,ms \\
    LLM only (no monitor) & 0/4    & OOM \\
    \bottomrule
  \end{tabular*}
\end{table}

\section{Prompt Engineering and Scheduler Example}
\label{app:prompt}

The LLM scheduler prompt evolved through six iterations.  The initial
version provided general contention guidance without explicit
comparison rules, causing the LLM to retain CPU-preferred tools on the
GPU.  The second version introduced hard threshold rules, which
improved accuracy but created contradictions near threshold
boundaries.  The third version removed thresholds in favor of direct
millisecond comparison, but still suffered from overly strong
``keep both on GPU'' guidance.  The fourth and fifth versions combined
explicit DAG-stage descriptions with structured per-tool reasoning
before the final JSON output, improving high-impact decisions in
Diamond DAG experiments.  The sixth version extended the prompt to the
three-way memory-constrained setting by describing
\texttt{gpu\_now}, \texttt{gpu\_queue}, and \texttt{cpu} factually and
integrating running averages, reprobe results, and swap-reprobe data
from the runtime monitor.  The fourth and fifth versions are used for
fan-out scenarios (S5--S10), where binary CPU/GPU placement benefits
from explicit ratio-based offload guidelines.  The sixth version is
used for three-way memory-constrained scenarios (S11--S13), where
structural memory and topology rules replace ratio thresholds because
the dominant constraint is memory feasibility rather than per-tool
speedup.

Figure~\ref{lst:prompt} shows an abbreviated prompt excerpt from
Round~3 of scenario S13.  After two rounds, the best observed mapping
is
$A{:}\texttt{gpu\_now},\,B{:}\texttt{gpu\_now},\,C{:}\texttt{gpu\_queue},\,D{:}\texttt{cpu}$
with $E{=}4581$\,ms.  The monitor's swap-reprobe block reports that
exchanging $B{\leftrightarrow}C$ yields $E{=}1585$\,ms at 1044\,MB,
within the 1100\,MB budget.  Figure~\ref{lst:cot} shows the
model-generated scheduling rationale and the final JSON mapping.  The
monitor supplies only measured values and feasibility outcomes; the
mapping decision is produced by the LLM scheduler.

\begin{figure}[t]
\small
\begin{tcolorbox}[colback=gray!5, colframe=gray!40, boxrule=0.5pt,
  left=4pt, right=4pt, top=3pt, bottom=3pt]
\begin{verbatim}
## Topology: Fan-out A||B||C||D.  VRAM=1100 MB.

Tool profiles (isolated):
  A clip_x12:        mem=627 gpu=147 cpu=3018 load=862
  B bert_base_x12:   mem=463 gpu=153 cpu=12141 load=245
  C whisper_x4:      mem=417 gpu=136 cpu=2156 load=1212
  D groupby_x60:     mem= 58 gpu= 90 cpu= 520 load= 202

Running averages (from observed rounds):
  A gpu_now=866 (2) | B gpu_now=571 (2) | C gpu_queue=3259 (1)
  D cpu=821 (2)

History (best-ever: round 2, E=4581, A:GN B:GN C:GQ D:CPU)
 0  A:GN B:GN C:GN D:CPU   FAIL (1508MB>1100MB)
 1  A:GN B:GN C:CPU D:CPU  E=4867 (1090MB) ok
*2  A:GN B:GN C:GQ  D:CPU  E=4581 (1090MB) ok

What-If (base=round 2, single-tool + swap reprobes):
  A->cpu         E=10836  (mem=464,  ok)
  B->cpu         E=28923  (mem=626,  ok)
  C->gpu_now     INFEASIBLE (mem=1508>1100)
  D->gpu_queue   E= 4146  (mem=1090, ok)
  SWAP C<->A     E= 2565  (mem= 882) full: A:GQ B:GN C:GN D:CPU
  SWAP C<->B     E= 1585  (mem=1044) full: A:GN B:GQ C:GN D:CPU
\end{verbatim}
\end{tcolorbox}
\caption{Abbreviated Round~3 prompt for scenario S13.
  GN = \texttt{gpu\_now}, GQ = \texttt{gpu\_queue}.  The starred
  history row is the best observed mapping.  The monitor reports
  measured end-to-end latency and memory usage for each probe; no
  recommendation is attached.  The VRAM values shown
  (\texttt{mem=627/463/417/58}) reflect peak measurements taken at
  experiment time; subsequent re-profiling with finer-grained peak
  tracking yields slightly higher values in the main paper's tool-profile
  table.  The LLM's reasoning is unaffected by this drift because it
  operates on the values it actually observes in the prompt.}
\label{lst:prompt}
\end{figure}

\begin{figure}[t]
\small
\begin{tcolorbox}[colback=blue!3, colframe=blue!30, boxrule=0.5pt,
  left=4pt, right=4pt, top=3pt, bottom=3pt]
\emph{The current mapping matches the best-ever and has an E2E of
4581\,ms.
According to the What-If analysis, swapping
C(whisper\_x4)$\leftrightarrow$B(bert\_base\_x12) results in a lower
E2E of 1585\,ms, which is a significant improvement.
This swap is within the memory budget.}

\medskip
\texttt{\{"A": "gpu\_now", "B": "gpu\_queue", "C": "gpu\_now", "D": "cpu"\}}
\end{tcolorbox}
\caption{Model-generated scheduling rationale for the prompt in
  Figure~\ref{lst:prompt} (Round~3 of S13).  The model quotes the
  monitor's measured end-to-end values, applies the memory-budget
  check, and emits the swapped mapping in JSON.  The adopted mapping
  matches the brute-force optimum and is retained for subsequent
  rounds.}
\label{lst:cot}
\end{figure}

\section{Complete S13 Mapping Trace}
\label{app:s13-trace}

\begin{figure*}[t]
  \centering
  \includegraphics[width=0.8\textwidth]{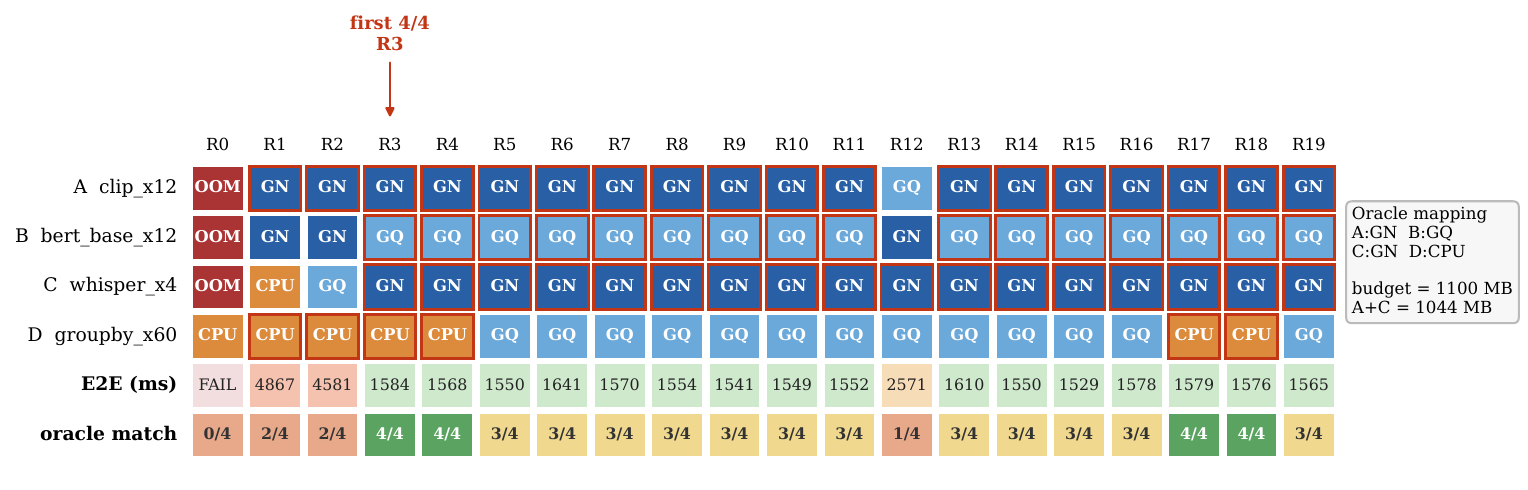}
  \caption{Complete S13 mapping trajectory over 20 rounds.  The main
    paper shows the event-level convergence storyboard; this appendix
    figure reports the full per-round mapping log.  Red outlines
    denote per-tool agreement with the oracle placement.}
  \label{fig:s13-trace}
\end{figure*}

Figure~\ref{fig:s13-trace} reports the per-round device mapping for the
full 20-round warm-start run of scenario S13.  Rows show the four
tools, the measured end-to-end latency, and the oracle-match count for
that round; cells matching the oracle placement are outlined in red.
R0 is memory-infeasible (three \texttt{gpu\_now} tools exceed the
1100\,MB budget); R1 and R2 are feasible but suboptimal; R3 is the
first 4/4 oracle match and the mapping adopted by the LLM converges to
\texttt{A:GN, B:GQ, C:GN, D:CPU}.  D continues to oscillate between
\texttt{cpu} and \texttt{gpu\_queue} in subsequent rounds with
negligible end-to-end impact, as discussed in the main-paper S13 case
study.

\section{Steady-State Latency Under Parallel Contention}
\label{app:steady-state}

The main-paper normalized-latency figure reports the best of 20
rounds for each adaptive strategy.  This minimum is subject to selection
variance (taking the lowest of 20 noisy measurements) and is the source
of the apparent ratios slightly below $1.0$ in that figure.
Table~\ref{tab:steady-state} reports a noise-resistant alternative
for the six fan-out scenarios S5--S10: mean and standard deviation
of measured E2E latency across all rounds in which the strategy
emitted its best-found mapping.

Across all six fan-out scenarios, UCB1, LLM-Cold, and LLM-Warm
converge to the brute-force oracle mapping; their steady-state means
agree with the oracle reference within $0.4$--$4\%$, well within
$1$--$2$ standard deviations of run-to-run noise.  The apparent
``below $1.0$'' ratios in the main-paper normalized-latency figure are
therefore single-round selection artifacts rather than improvements over
the oracle.  StarPU's online performance model converges to the all-GPU
mapping in S5--S6 and S8 and pays the full contention penalty
(\textgreater\,$1.3\times$ over oracle); for S7, S9, and S10 it
matches the oracle within run-to-run noise.

\begin{table*}[t]
  \caption{Fan-out steady-state E2E latency (ms) by strategy.  Each
    cell reports $\mu \pm \sigma$ measured across rounds emitting the
    strategy's best-found mapping (sample size $n$ shown in
    parentheses; range across cells is $n\!=\!4$ to $20$).  Oracle
    column is a single brute-force measurement of the oracle
    mapping.  Cells whose mapping equals the oracle mapping are
    bold.}
  \label{tab:steady-state}
  \small
  \centering
  \begin{tabular*}{\textwidth}{@{\extracolsep{\fill}}lrcccc@{}}
    \toprule
    Scenario & Oracle (BF) & UCB1 & StarPU & LLM-Cold & LLM-Warm \\
    \midrule
    S5  & 599.4 & \textbf{599.2 $\pm$ 2.2} (5)  & 792.4 $\pm$ 3.1 (20) & \textbf{600.5 $\pm$ 2.6} (4)  & \textbf{599.7 $\pm$ 3.6} (5)  \\
    S6  & 508.3 & \textbf{508.4 $\pm$ 3.8} (5)  & 712.4 $\pm$ 1.6 (20) & \textbf{506.3 $\pm$ 3.1} (6)  & \textbf{507.2 $\pm$ 3.0} (8)  \\
    S7  & 655.4 & \textbf{658.6 $\pm$ 0.8} (5)  & \textbf{659.7 $\pm$ 1.5} (20) & \textbf{658.3 $\pm$ 1.0} (7)  & \textbf{658.3 $\pm$ 0.8} (7)  \\
    S8  & 492.1 & \textbf{497.5 $\pm$ 5.9} (5)  & 706.9 $\pm$ 2.3 (20) & \textbf{499.1 $\pm$ 4.7} (4)  & \textbf{496.7 $\pm$ 3.3} (8)  \\
    S9  & 439.8 & \textbf{440.9 $\pm$ 3.8} (5)  & \textbf{442.9 $\pm$ 3.0} (20) & \textbf{439.9 $\pm$ 3.0} (13) & \textbf{442.9 $\pm$ 1.6} (6)  \\
    S10 & 649.7 & \textbf{646.8 $\pm$ 1.5} (5)  & \textbf{648.8 $\pm$ 1.9} (20) & \textbf{648.1 $\pm$ 3.2} (12) & \textbf{646.7 $\pm$ 2.3} (6)  \\
    \bottomrule
  \end{tabular*}
\end{table*}

\end{document}